\documentclass[12pt,a4paper]{article}
\usepackage{amsmath}
\usepackage{graphicx}
\usepackage{amsfonts}
\usepackage{amssymb}
\usepackage{epsfig}
\usepackage{psfrag}
\usepackage{wrapfig}
\usepackage{color}

\definecolor{yellow}{rgb}{1,0.9,0} 
\definecolor{wine}{rgb}{0.5,0,0.4}
\definecolor{lightblue}{rgb}{0.3,0,0.7}
\definecolor{ancientrose}{rgb}{0.7,0,0.4}
\definecolor{cream}{rgb}{1.,1.,0.7}
\definecolor{violet}{rgb}{1.,0.9,0.95}
\definecolor{lightgreen}{rgb}{0.8,1,0.8}
\definecolor{darkgreen}{rgb}{0,0.6,0}
\newcommand{\be}{\begin{equation}}
\newcommand{\ee}{\end{equation}}
\newcommand{\beq}{\begin{eqnarray}}
\newcommand{\eeq}{\end{eqnarray}}

\def\numu{\mathrel{{\nu_\mu}}}
\def\nutau{\mathrel{{\nu_\tau}}}

\def\barnue{\mathrel{{\bar \nu}_e}}

\def\barnux{\mathrel{{\bar \nu}_x}}

\def \lta {\mathrel{\vcenter{\hbox{$<$}\nointerlineskip\hbox{$\sim$}}}}
\def \gta {\mathrel{\vcenter{\hbox{$>$}\nointerlineskip\hbox{$\sim$}}}}

\def\t13{\mathrel{{\sin^2 \theta_{13}}}}
\def\y12{\mathrel{{\tan^2 \theta_{12}}}}

\begin{document}

\begin{titlepage}

\vspace*{0.2cm}

\begin{center}


{\Large \bf Neutrinos from SN1987A: flavor conversion and interpretation of results
 } \\

\vspace{0.8cm}
{\large
Cecilia Lunardini$^{a,1}$ and
Alexei Yu. Smirnov$^{b,c,2}$ } \\

\vspace{0.2cm}
{\em $^{a}$ Institute for Advanced Study, Einstein drive, 08540 Princeton,
New Jersey, USA}

\vspace{0.2cm}
{\em $^{b}$ The Abdus Salam ICTP, Strada Costiera 11, 34100 Trieste, Italy}

\vspace{0.2cm}
{\em $^c$ Institute for Nuclear Research, RAS, Moscow 123182, Russia.}

\end{center}
\begin{abstract}
After recent results from solar neutrino experiments and KamLAND we
can definitely say that neutrinos from SN1987A underwent flavor
conversion, and the conversion effects must be taken into account in
the analysis of the data.  Assuming the normal mass hierarchy of
neutrinos we calculate the permutation factors $p$ for the
Kamiokande-2, IMB and Baksan detectors.  The conversion inside the
star leads to $p = 0.28 - 0.32$; the oscillations in the matter of the
Earth give partial (and different for different detectors)
regeneration of the original $\bar{\nu}_e$ signal, reducing this
factor down to 0.15 - 0.20 (at $E = 40$ MeV).
We study in details the influence of conversion on the observed signal
depending on the parameters of the original neutrino spectra.  For a
given set of these parameters, the conversion could lead to an
increase of the average energy of the observed events up to 50\% and
of the number of events by a factor of 2 at Kamiokande-2 and by a
factor of 3 - 5 at IMB. Inversely, we find that neglecting the
conversion effects can lead up to 50\% error in the determination of
the average energy of the original $\bar{\nu}_e$ spectrum and about
50\% error in the original luminosity.  Comparing our calculations
with experimental data we conclude that the Kamiokande-2 data alone do
not favor strong conversion effect, which testifies for small
difference of the original $\bar{\nu}_e$ and $\bar{\nu}_\mu$ spectra.
In contrast, the combined analysis of the Kamiokande and IMB results
slightly favors strong conversion effects (that is, large difference
of the original spectra). In comparison with the no oscillation case,
the latter requires lower average energy and higher luminosity of the
original $\bar{\nu}_e$ flux. 

\end{abstract}

\vfil
\noindent
\footnoterule
{\small
$^{1}$ E-mail: lunardi@ias.edu\vskip-1pt\noindent}
{\small $^2$ E-mail: smirnov@ictp.trieste.it\vskip-1pt\noindent}

\thispagestyle{empty}
\end{titlepage}

\setcounter{page}{1}

\section{Introduction}

After recent results from solar neutrino experiments, and first of all
SNO~\cite{Ahmad:2001an,Ahmad:2002jz,Ahmad:2002ka}, as well as from the
reactor experiment KamLAND~\cite{Eguchi:2002dm}, which have selected
the Large Mixing Angle (LMA) MSW solution of the solar neutrino
problem, we can definitely say that neutrinos from SN1987A got
converted. The neutrino flavor transformations influenced the signals
observed in 1987 by Kamiokande-2
(K2)\cite{Hirata:1987hu,Hirata:1988ad}, IMB
\cite{Bionta:1987qt,Bratton:1988ww} and Baksan
\cite{Alekseev:1987ej}. Conversion effects must be taken into account
in the analysis of the data and in the determination of the properties of the
original neutrino fluxes.  Results obtained without conversion are to
some extent incorrect.

The conversion of neutrinos associated to  SN1987A has
been extensively studied  before (see
\cite{Wolfenstein:1987pj}-\cite{Barger:2002px} for an incomplete list
of references). Here we comment on some papers which are relevant for
our present discussion.

It was suggested~\cite{talkalexei} that the difference of signals
detected by the K2 and IMB detectors can be explained partially by the
oscillations of neutrinos in the matter of the Earth since the
distances crossed by neutrinos on the way to these two detectors were
different. The suggestion implied, however, a large lepton mixing,
which was  not  a favored idea at that time. The detailed calculations
have been done 13 years later~\cite{Lunardini:2000sw}, when certain
hints appeared that LMA could be the correct solution of the solar
neutrino problem.

In connection to SN1987A, Wolfenstein considered antineutrino
conversion in the star in the case of large
mixing~\cite{Wolfenstein:1987pj}. He concluded that conversion leads
to a harder energy spectrum of the observed events and, possibly, to a
larger number of events.

In the attempt to restrict the large lepton mixing, the conversion of
antineutrinos in the non-resonance channel has been considered in
details~\cite{Smirnov:1994ku}. From the analysis of the SN1987A data
a bound on the permutation factor ($p < 0.35$), and consequently,on the 
mixing angle has been obtained. It was found that the bound is weaker
in the LMA range, and the Earth matter effect further relaxes it.

A detailed analysis of the SN1987A data based on the Poisson statistics
has been performed by Jegerlehner et al. \cite{Jegerlehner:1996kx},
who found that the data do not allow a definite conclusion on
the oscillations hypothesis.  In the event that a large neutrino
mixing is confirmed (as it has been recently), the data analysis would
point toward average neutrino energies (at production in the star)
lower than what theoretically predicted.  By combining solar and
SN1987A data, the authors of
refs. \cite{Kachelriess:2001sg,Kachelriess:2000fe} concluded that the
LMA region was the most suitable, among the large mixing solutions of
the solar neutrino problem, to reconcile SN1987A data and predictions
from numerical supernova codes, in agreement with \cite{Smirnov:1994ku}.

In ref. \cite{Lunardini:2000sw}, following the early
suggestion~\cite{talkalexei}, we have considered the possibility that
certain features of the energy spectra of the events detected by K2
and IMB can be explained by neutrino oscillations in the matter of the
Earth. This fixes several bands in the $\Delta m^2 - \cos 2\theta$
plane. It was concluded that the data favor the parameters of LMA solution
and the normal mass hierarchy.  The inverted mass hierarchy is
disfavored, unless the $1-3$ mixing angle is very small
\cite{Minakata:2000rx} (see however \cite{Barger:2002px}).

The combined analysis of results from solar neutrino experiments and
KamLAND lead to the values of oscillation
parameters 
\be 
\Delta m^2 = 7.1^{+3.2}_{-2.2} \cdot 10^{-5} {\rm eV}^2
, ~~~ \tan^2 \theta = 0.40 \pm 0.10~,
\label{bfglob} 
\ee which coincides with the third band (from the bottom in $\Delta m^2$
scale) found in~\cite{Lunardini:2000sw}.

In this paper we revisit the conversion of neutrinos from SN1987A using the
latest information on neutrino mass spectrum and mixing.  We address the
questions of how neutrinos were converted, how  conversion modified  the
observed signals and what could be the error in the determination of the
original neutrino fluxes if conversion is not taken into account.\\ 

The analysis of the SN1987A data and a determination of the original
 spectra as precise as possible are needed not only to understand what
 happened in 1987 but also to compare with the results of future
 detections of neutrino bursts from supernovae. Detections of
 supernova neutrinos are rare events and furthermore each supernova is
 unique. Indeed, the mass of the progenitor, luminosity, rotation,
 magnetic fields, chemical composition can be substantially different,
 and, as a consequence, the properties of the neutrino fluxes
can vary. Thus,
future high statistics detections are not expected to reproduce the same features as those
of SN1987A, but will give somehow complementary information. 
The comparison of neutrino signals from different supernovae would be
extremely important for understanding the latest stage of star
evolution, the dynamics of core collapse and explosion.\\

The paper is organized as follows. In sec. 2 we consider conversion of
antineutrinos and calculate the $\bar{\nu}_e$ survival probabilities
and permutation factors for Kamiokande-2, IMB and Baksan. In sec. 3 the
effects of conversion on the observed signals are studied depending on
the parameters of the original spectra. In sec. 4 we compare the
predictions with the experimental results and make some indicative
conclusions on the properties of the original fluxes. The results are
summarized in sec. 5.

\section{ Neutrino conversion. Permutation factor}

As a consequence of the equality of the original $\bar{\nu}_{\mu}$ and
$\bar{\nu}_{\tau}$ fluxes, $F_{\bar\mu}^0 = F_{\bar \tau}^0 \equiv F_{
\bar x}^0$, the conversion effects in the antineutrino channel are
described by the unique probability $P_{ee} \equiv P(\bar{\nu}_e
\rightarrow \bar{\nu}_e)$ which is the total $\bar{\nu}_e$ survival
probability from the production point to the
detector~\cite{Dighe:1999bi} (see
however~\cite{Akhmedov:2002zj}). $P_{ee}$ takes into account the
conversion/oscillation effects inside the star, on the way from the
star to the Earth and oscillations in the matter of the Earth. Using
$P_{ee}$, the electron antineutrino flux at the detector, $F_{\bar e}$,
can be written in terms of the original fluxes as 
\be F_{\bar e} =
F_{\bar e}^0 + (1 - P_{ee}) \Delta F^0, \label{fluxe} 
\ee where 
\be
\Delta F^0 \equiv F_{\bar x}^0 - F_{\bar e}^0 
\label{delta} 
\ee is the
difference of original fluxes. The combination \be p \equiv 1- P_{ee}
\ee is often called the permutation factor (indeed, if $P_{ee} = 0$,
$p = 1$ and $F_{\bar e} = F_{\bar x}^0$, that is, as a result of
conversion the initial spectra are permuted).

After the recent determination of the oscillation parameters,  and with a
few plausible  assumptions, the  physics of  conversion  is basically
determined and the probability $P_{ee} = P_{ee}(E)$ can be  calculated reliably.

In general, taking into account the loss of coherence between mass eigenstates
on the way from the star to the Earth, we can write \be P_{ee} = \sum_i
P_{ei}^{SN} \times P_{ie}^{Earth}, \label{probab} \ee where $P_{ei}^{SN}$ is the
probability of the $\bar{\nu}_e \rightarrow \bar{\nu}_i$ transition inside the
star, and $P_{ie}^{Earth}$ is the probability of $\bar{\nu}_i \rightarrow
\bar{\nu}_e$ transition inside the Earth. Here $\bar{\nu}_i$ are the 
neutrino mass eigenstates.

We assume the following. 

\begin{itemize}

\item
There are only three neutrinos, and if additional (sterile) neutrinos
exist, their mixings with active neutrinos are  negligible.

\item
Neutrinos  have normal mass hierarchy, or if the hierarchy is inverted, the
1-3 mixing is very small, so that its effect can be neglected (see \cite{Dighe:1999bi}).

\item 
The density profile in the region of the 1-2 resonance coincides with the
progenitor profile during the neutrino burst ($\sim 10$ s).  It is not
affected by shock wave propagation.  Even if the profile changes with
time, the adiabaticity character of the conversion remains unchanged due
to the large 1-2 mixing.

\end{itemize}

(We will comment  on  effects of possible relaxation of these conditions in the
sec. 5.)

Under the above assumptions the dynamics of neutrino conversion is completely
determined:

1). Inside the star   the electron antineutrinos are adiabatically converted
into the mass eigenstate $\bar{\nu}_1$~\cite{Dighe:1999bi} 
\be 
\bar{\nu}_e
\rightarrow \bar{\nu}_1, 
\label{instar} 
\ee 
and consequently, $P_{e1}^{SN} = 1$,
whereas $P_{ei}^{SN} = 0$ for $i \neq 1$. Indeed, in 
the assumption of normal mass hierarchy there are no level crossings 
in the antineutrino channel. 
Furthermore, in the production point the mixing is strongly suppressed, so that there we have
$\bar{\nu}_e = \bar{\nu}_{1m}$, where $\bar{\nu}_{1m}$ is the eigenstate of the
neutrino propagation in matter. For the parameters (\ref{bfglob}) the
propagation is highly adiabatic 
and so the neutrino state  coincides with $\bar{\nu}_{1m}$ during 
the whole propagation.   
At the surface of the star (zero matter density) $\bar{\nu}_{1m}$ 
equals $\bar{\nu}_{1}$. Corrections due to possible deviations from the
adiabaticity are negligible. 

2). Being an eigenstate of the Hamiltonian in vacuum, $\bar{\nu}_{1}$
propagates without any change from the surface of the star to the
Earth.

3). Inside the matter of the Earth $\bar{\nu}_{1}$ oscillates, in particular to
$\bar{\nu}_e$: 
\be 
\bar{\nu}_1 \rightarrow \bar{\nu}_e. 
\label{inearth} 
\ee 
We denote the probability of this transition as  $P_{1e}^{Earth}$.\\ 

According to this physical picture and eq. (\ref{probab}), the survival probability can be
written immediately: 
\be P_{ee} = P_{1e}^{Earth}. 
\label{probearth} 
\ee 
Thus,  the total survival $\bar{\nu}_e
\rightarrow \bar{\nu}_e$ probability simply coincides with $P_{1e}^{Earth}$
inside the Earth. If the star is not shielded by the Earth, we find 
\be
P_{1e}^{Earth} = |U_{e1}|^2 \equiv \cos^2\theta_{13} \cos^2 \theta_{12}
\label{vac} 
\ee 
in the standard parameterization of the vacuum mixing matrix (see
{\it e.g.} \cite{Krastev:1988yu}). Furthermore, if the 1-3 mixing is small
enough the probability equals 
\be 
P_{ee} \approx \cos^2 \theta_{12}~,
\ee 
where the angle $\theta_{12}$ is immediately obtained from the
analysis of the solar neutrino data and KamLAND results in the two
neutrino framework, eq.  (\ref{bfglob}).

If the 1-3 mixing is near the present upper bound, $\sin^2
\theta_{13}\lta 0.06$
\cite{Apollonio:1998xe,Apollonio:1999ae,Boehm:2000vp,Boehm:2001ik},
and we are interested in $\sim 10\%$ corrections, the complete
expression (\ref{vac}) should be used. The solar neutrino data allow
us to measure immediately the combination $\cos^4 \theta_{13} \sin^2
\theta_{12}$ in high energy experiments and $\cos^4 \theta_{13} (1 -
0.5 \sin^2 2\theta_{12})$ in future low energy experiments. These
combinations differ from that in eq. (\ref{vac}), therefore
independent measurements of $\theta_{13}$
are needed to know $P^{Earth}_{1e}$ accurately. In the present
discussion of the SN1987A signal we can neglect corrections due to
non-zero 1-3 mixing, since the errors produced in doing so are within
the present accuracy of determination of $\theta_{12}$. \\

The  probability of oscillations inside the Earth can be written as 
\be
P_{1e}^{Earth} \equiv |U_{e1}|^2 + \bar{f}_{reg}, 
\label{mat} 
\ee 
where
$\bar{f}_{reg}$ is called the $\bar{\nu}_e$ regeneration factor.  Notice
that $\bar{f}_{reg} > 0$, because inside the Earth the  matter density is
higher than in the cosmic space.  
So,   there is a partial return to the initial condition of high density
in the production point, where the neutrino state is $\bar{\nu}_e$. Thus, the
oscillations inside the Earth weaken the net conversion effect.

In the two neutrino context, the permutation factor can be written then as 
\be 
p = \sin^2\theta_{12} - \bar{f}_{reg}, 
\label{perreg}
\ee 
and  the  flux of the electron
antineutrinos at the detector equals: 
\be 
F_{\bar e} = F_{\bar e}^0 + p \Delta F^0
\approx F_{\bar e}^0 + (\sin^2\theta_{12} - \bar{f}_{reg})\Delta F^0.
\label{fluxeearth} 
\ee 
In spite of its simplicity, this expression gives a precise
description of the neutrino conversion effect. The permutation
factor does not depend on time during the neutrino burst, in the
assumption that the matter density profile in the star does not undergo a
drastic time evolution.\\

\begin{figure}[hbt]
\begin{center}
 \epsfig{file=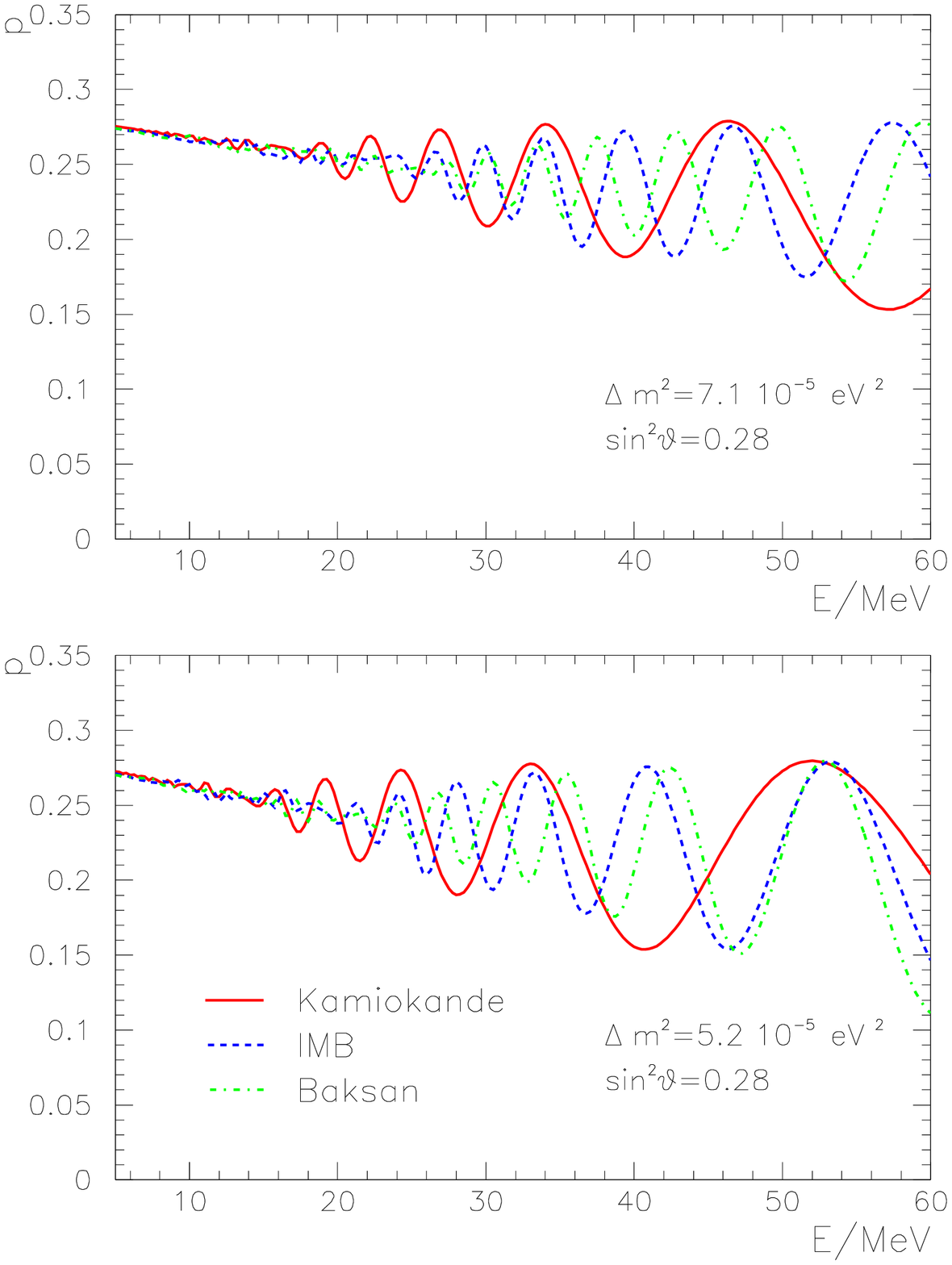,
width=13truecm}
\end{center}
\caption{The permutation factor, $p=1-P_{1e}$, as a function of the
neutrino energy at Kamiokande-2, IMB and Baksan. The two panels refer
to two choices of the oscillation parameters from the LMA allowed
region, one of which (upper panel) is the best fit set of values in
eq. (\ref{bfglob}). For illustration purposes, we have averaged $p$
over energy according to eq. (\ref{eq:av}), with $\Delta E=2$ MeV.}
\label{fig:1}
\end{figure}

We have performed exact numerical calculations of the probability
$P_{1e}^{Earth}$ and the regeneration factor using the Earth density
profile from~\cite{Dziewonski:1981xy}. The fig. \ref{fig:1} shows the
permutation factor $p = 1-P_{1 \bar e}^{Earth}$ for Kamiokande-2, IMB
and Baksan detectors for two different choices of the oscillation
parameters from the LMA allowed region. To remove the unobservable
fast oscillations we have averaged the probability over energy as
follows: 
\be 
\langle P(E) \rangle = \frac{1}{\Delta E} \int^{E+\Delta
E/2}_{E-\Delta E/2} P(E') dE' ~. 
\label{eq:av} 
\ee In the figures
$\Delta E=2$ MeV was taken for illustration.  \\ 
The following remarks are in order.

Due to the larger distance crossed by neutrinos inside the Earth, for the  
IMB detector the frequency
of the oscillatory  curve in the energy scale is  twice as
large as the frequency for K2. The depth of oscillations is nearly the same 
for all detectors, but the phase of oscillations is different.

Also the averaged permutation factor $\bar p$ 
has the same energy dependence for all detectors: 
For the best fit values of oscillation parameters (upper panel)
it decreases from 0.28 down to 0.25 at 
$E =  25$ MeV and 0.23 at $E =  40$ MeV. 

For K2, at low energies ($E < 25$ MeV), the oscillations in the matter of the Earth 
are essentially averaged out.  
For higher energies the averaging is not complete. 
Notice that the minima of the oscillatory curve are 
at $E = 30$ ($p = 0.21$ ) and 40 MeV ($p = 0.19$). 
For IMB the minima are at 36 MeV and 43 MeV ($p = 0.19$).

For comparison we show also the permutation factors for 
$\Delta m^2 = 5.2 \cdot 10^{-5}$ eV$^2$ which is at the border of the 
allowed region and in the lower band identified in \cite{Lunardini:2000sw} from the 
spectral features. 

Notice that with the decrease of $\Delta m^2$ both the depth of modulation
and its period increase. In particular, $f_{reg} \propto 1/\Delta
m^2$.  Now in the minimum of the K2 curve at 40 MeV we find $p =
0.15$.  The minima are also slightly shifted in energy.

The permutation factor increases with the mixing according to
eq. (\ref{perreg}); at the same time the regeneration factor changes
weakly.

\begin{figure}[hbt]
\begin{center}
 \epsfig{file=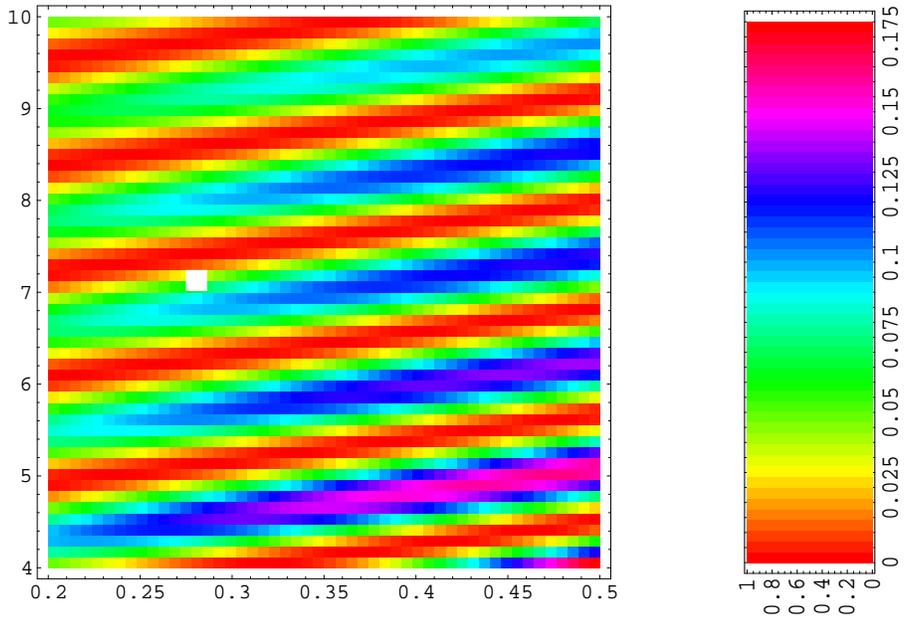,
width=12truecm}
\end{center}
\caption{Regions of constant values of the regeneration factor $\bar f_{reg}$ at K2
and IMB in the plane $\Delta m^2/(10^{-5}~{\rm eV^2})-\sin^2 \theta$
(vertical and horizontal axes respectively), for neutrino energy
$E=40$ MeV. The white squares denote the LMA  best fit point, eq. (\ref{bfglob}). } \label{fig:lma}
\end{figure}

The fig. \ref{fig:lma} shows values of  the regeneration factor $\bar f_{reg}$
for K2 and IMB in the $\Delta m^2 - \sin^2 \theta_{12}$  plane for $E=40$ MeV.  
For other energies the result can be
obtained from fig. \ref{fig:lma} by rescaling of $\Delta m^2$: $\bar{f}_{reg}(E,
\Delta m^2) = \bar{f}_{reg}(\Delta m^2 \times (40{\rm MeV}/E))$. 
For the best fit values of the oscillation parameters
we have  $\bar{f}_{reg} \simeq  0.08$ at Kamiokande-2 for E = 40 MeV.\\

Concluding this part we can definitely say, that

\begin{itemize}

\item
the permutation factor  due to conversion inside the star is about 0.2 - 0.4
with most plausible value 0.28 - 0.30;

\item for all detectors this factor is modulated significantly by the Earth
matter effect, which suppresses the permutation down to $ \sim 0.17$ for E = 40
MeV. The modulation effect  due to oscillations inside the Earth increases with energy.

\end{itemize}

The effect inside the star is determined mainly by the mixing $\sin^2
\theta_{12}$ and practically does not depend on $\Delta m^2$. In
contrast, the oscillations inside the Earth are very sensitive to
$\Delta m^2$ and also depend on mixing. Future operation of the
KamLAND experiment will allow us to determine $\Delta m^2$ even
better, and therefore to have more precise determination of the Earth
matter effect. Improvements in the determination of the mixing angle
may follow from further operation of SNO.

Notice that in spite of the relatively  small values of the permutation factor
and even smaller Earth matter effect, the  influence of  conversion on the
neutrino fluxes can be strong. Indeed, if the original $\bar\nu_{x}$ flux has
substantially larger average energy than the $\bar\nu_{e}$ flux, then 
$F_{\bar e}^0 \ll F_{\bar x}^0$ in the high energy tail of
the spectrum, due to the exponential decrease of the fluxes with the increase of
energy. Therefore the observed flux at high energy will be dominated by
the converted flux (second term in eq. (\ref{fluxe})).

\section{Conversion effects and  original spectra}
\label{sec3} 

According to eq. (\ref{fluxe}) the effect of conversion 
on the neutrino spectra, and, consequently, 
on the spectra of observed events, is proportional to
the difference of the original $\barnue$ and $\barnux$ fluxes. 
In view of the large uncertainties in these 
original fluxes (see e.g. the summary in \cite{Keil:2002in}) we will
study systematically the conversion effects depending 
on the values of the parameters of the original  spectra.

We will use the following parametrization of  the instantaneous original fluxes
of neutrinos, which depends explicitly on the integral
characteristics~\cite{Keil:2002in}: 
\be 
F^0_z (E) = \frac{L_z}{4 \pi D^2
E^2_{0z}} \frac{(1+\alpha_z)^{1+\alpha_z}}{\Gamma(1+\alpha_z)}
\left(\frac{E}{E_{0z}} \right)^{\alpha_z} \exp \left[ -(1+\alpha_z) E/E_{0z}
\right]~, 
\label{orig} 
\ee 
where $L_z$ is the total (integrated over the
neutrino energy) luminosity in the flavor $z$ and $E_{0z}$ is the average energy
(as it can be checked by explicit calculation); $D$ is the distance to the star;
and $\alpha_z$ plays the role of a pinching parameter.  It  is estimated
to be in the interval ~\cite{Keil:2002in} 
\be \alpha_z\sim 2 - 5~. 
\label{alp}
\ee 
The width of the spectrum (\ref{orig}) equals 
\be 
\gamma(\alpha_z) \equiv
\left[\frac{\langle E^2 \rangle}{\langle E \rangle^2}-1 \right]^{1/2} =
\frac{1}{\sqrt{1+\alpha_z}}~. 
\label{widFD} 
\ee
The Fermi-Dirac spectrum corresponds to $\alpha_z \simeq 2.3$. 

The parameters of neutrino fluxes  change with time during the neutrino burst:
$E_{0z} = E_{0z}(t)$, $L_z = L_z(t)$, $\alpha_z = \alpha_z(t)$. In particular,
the spectra  may have two different time components~(see e.g. the review in
\cite{Loredo:2001rx} and references therein):  one from the accretion phase,
and another one from the cooling phase. During these phases the parameters of the 
spectra can be substantially different whereas within each phase they change
slowly.

The integral characteristics of the original spectra relevant for our discussion
are the average energy of the electron antineutrino spectrum, $E_{0e}$; the
luminosity in the electron antineutrinos, $L_e$; the ratio of the  average
energies of the muon/tau and electron antineutrino spectra, $r_E$, and the ratio
of the corresponding luminosities,  $r_L$: 
\beq r_E \equiv
\frac{E_{0x}}{E_{0e}}, ~~~~r_L \equiv \frac{L_x}{L_e}~;  
\eeq 
the pinchings of the $\bar{\nu}_e$  and  $\bar{\nu}_x$ spectra,  $\alpha_e$ and $\alpha_x$. \\

In the following we consider the signals observed in the Cerenkov
detectors K2 and IMB. The features of the predicted neutrino signal at
Baksan and the consistency with the observed data are discussed at the
end of sec.  \ref{sec4}.  The differential spectra of the detected
positrons from $\barnue + p \rightarrow e^+ + n$ are given by
\begin{eqnarray}
\frac{dN^{i}}{d\epsilon} =
N^{i}_T\int_{-\infty}^{+\infty} d\epsilon' {\cal R}^{i}(\epsilon,\epsilon')
{\cal E}^{i}(\epsilon') \int dE F_e^{i}(E)
\frac{d\sigma (\epsilon', E) }{ d\epsilon'}~,
\label{eq:dnem}
\end{eqnarray}
where $i = {\rm K2, IMB}$, $\epsilon$ and $\epsilon'$ are
the observed and the true energies of the positron respectively, $N_T^{i}$ is
the number of target particles in the fiducial volume, ${\cal E}^{i}$ is the
detection efficiency. In (\ref{eq:dnem}) the energy resolution function, ${\cal
R}^{i}(\epsilon, \epsilon')$, is taken in the Gaussian form with
$\sigma_{\epsilon} = 0.87 \sqrt{\epsilon/{\rm MeV}}$ for K2
\cite{Hirata:1988ad} and $\sigma_{\epsilon } = 1.16 \sqrt{\epsilon /{\rm
MeV}}$ for IMB \cite{Bionta:1987qt}. The fluxes of antineutrinos at the
detectors, $F^{i}_e(E)$ are given by  eqs.
(\ref{orig}) and (\ref{fluxeearth}). We use the differential cross section of
the detection reaction, ${d\sigma (\epsilon ', E)/d\epsilon '}$, calculated
in~\cite{Strumia:2003zx}. 
The conversion effects ($p^{i}$) have been found using the best fit
values of oscillation  parameters given in eq. (\ref{bfglob}).\\

The  signals at the various detectors will be
characterized by the following integral characteristics.  

- rate, or total number of the observed events $N^{i}$:
\beq
N^{i} = \int_{\epsilon_{th}^{i}}^{+\infty} d\epsilon
\frac{dN^{i}}{d\epsilon};
\label{number}
\eeq

- the average energy of the detected (electron antineutrino) events,
$\bar{\epsilon} ^{i}$: 
\beq 
\bar{\epsilon}^{i} = \frac{1}{N^{i}}
\int_{\epsilon_{th}^{i}}^{+\infty} d\epsilon~ \epsilon
\frac{dN^{i}}{d\epsilon}; 
\label{avav} 
\eeq

- width of the observed spectrum: 
\beq 
\Gamma^{i}  =\left[ \frac{\langle
\epsilon^2 \rangle^{i}}{(\bar{\epsilon}^{i})^2} -1 \right]^{1/2}, 
\eeq
where 
\beq 
\langle \epsilon^2 \rangle^{i} = \frac{1}{N^{i}}
\int_{\epsilon_{th}^{i}}^{+\infty} d\epsilon~ \epsilon^2
\frac{dN^{i}}{d\epsilon}. 
\label{widwid} 
\eeq

We calculate these characteristics of the K2 and IMB signals as functions
of  the parameters of the original neutrino spectra. 
The results can be considered as instantaneous characteristics of observed events or 
as the integral characteristics if the parameters of the original spectra do not change 
during the neutrino burst. 
This matches the constant temperature model presented in the upper panel of
Table 4 in ref. \cite{Loredo:2001rx} \footnote{ Strictly speaking,
the results in \cite{Loredo:2001rx} are not directly comparable to ours, 
since they were obtained using a different parametrization of the neutrino
original fluxes.}.

\begin{figure}[hbt]
\begin{center}
\epsfig{file=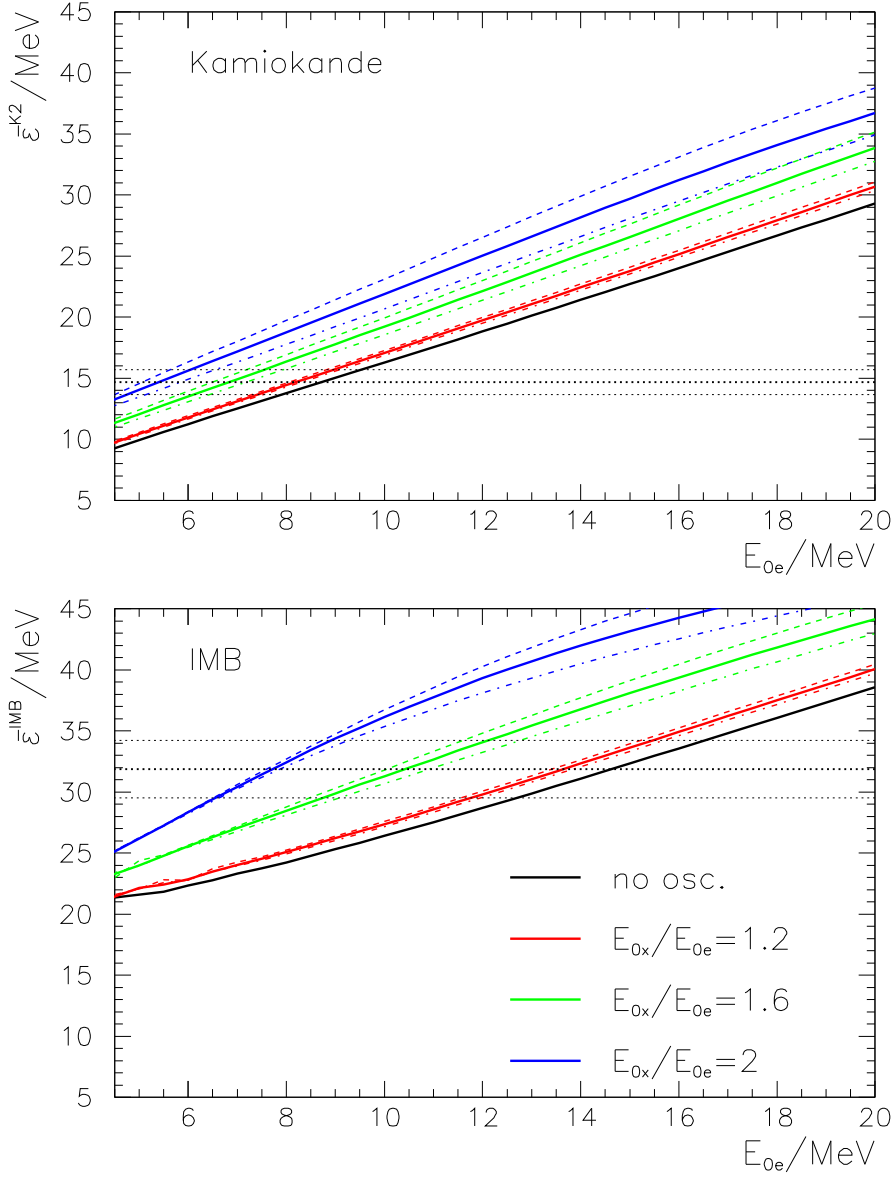, width=13truecm}
\end{center}
\caption{The average energy of positrons in the detectors K2 and IMB, $\bar
\epsilon^{i} $, as a function of the average energy of the original $\bar
\nu_e$ flux, $E_{0e}$, for the best fit LMA parameters (eq. (\ref{bfglob}), see
also fig. \ref{fig:1}) and different values of  $r_E=E_{0x}/E_{0e}$
(see legend) and $r_L=L_x/L_e$. The solid, dashed and dotted-dashed lines
correspond to $r_L=1,1.5,0.667$ respectively. The no oscillation case (or,
equivalently, the case $r_E=1$) is shown for comparison. The dotted (horizontal)
lines represent the values of  $\bar \epsilon^{i} $ extracted from the data
for each experiment, with the $1\sigma$ error (eq. (\ref{epsob})).}
\label{figure1.eps}
\end{figure}

Results for the observed numbers of events and average energies are
presented in the figs.  \ref{figure1.eps}-\ref{figure4.eps}, and
summarized in the Table \ref{tab:1}.
They refer to the case in which $\alpha_e=\alpha_x=3$, so that the
corresponding widths are equal and fixed to the value $\gamma=0.5$,
according to eq. (\ref{widFD}). Fig. \ref{figure7.eps} shows results
for the widths  $\Gamma^{i}$ and illustrates their dependence  on
$\alpha=\alpha_e=\alpha_x$.  A further illustration of the $\alpha$-dependence is given in  Table \ref{tab:2}.\\
\\

\begin{table}
\begin{center}
\begin{tabular}{lcccc}
\hline
         &  no oscill.
         &  $r_E=1.2$        &  $r_E=1.6$        & $r_E=2$
         \\
         \hline
         \hline
$\bar \epsilon^{K2}/\bar \epsilon_{no~ osc.}^{K2}-1$   & 0& 0.04-0.06 & 0.18 & 0.35 \\ 
$\bar \epsilon^{IMB}/\bar \epsilon_{no~ osc.}^{IMB}-1$   & 0& 0.04 & 0.18 & 0.35-0.37 \\ 
\hline 
$N^{K2}/N_{no ~osc.}^{K2}-1$  & 0& 0.08&  0.19 & 0.29 \\ 
$N^{IMB}/N_{no ~osc.}^{IMB}-1$ & 0 & 0.27&  0.36& 1.64 \\ 
\hline 
$E_{0e}/MeV$ (from $\bar \epsilon^{K2}$) & 8.8& 8.2& 6.7 & 5.4\\ 
$E_{0e}/MeV$  (from $\bar \epsilon^{IMB}$) & 14.7& 13.7& 10.4 & 7.6 \\ 
\hline
$L_e/L_0$  ($N^{K2} = 12$)   & 1& 0.93& 0.78 & 0.76\\
$L_e/L_0$  ($N^{IMB} = 8$)    & 2.75& 2.17&  1.2& 0.76\\ 
\hline
$\bar \epsilon^{K2}/\bar \epsilon^{IMB}$ 
& 0.65& 0.64& 0.63& 0.62\\
\hline
$N^{K2}/N^{IMB} $ 
& 4.2  &3.5  & 2.2 & 1.5 \\
\hline 
\hline
\end{tabular}
\caption{Effect of the neutrino conversion on characteristics of
observable K2 and IMB signals and extracted parameters of the original
neutrino spectra.  The numbers of events and luminosities 
correspond to $E_{0e}=11$ MeV and $r_L=1$.}
\label{tab:1}
\end{center}
\end{table}

Let us summarize the effects of flavor conversion:
\\

1). The conversion leads to an increase of the average energy of the observed
events (fig.~\ref{figure1.eps}). 
According to fig.~\ref{figure1.eps}, for K2   the dependence of $\bar{\epsilon} ^{i}$ 
on  $E_{0e}$ is linear  in a wide interval of
energies ($E_{0e} > 5$ MeV) due to the low energy threshold, and it can be
approximated with good accuracy by 
\be 
\bar{\epsilon} \approx A(r_E, r_L)
E_{0e} + B(r_E, r_L). 
\label{avE} 
\ee  
We find $A(r_E, 1) = 1.24,~ 1.32,~ 1.47,~  1.58$ for  $r_E = 1.0, ~ 1.2,~ 1.6,~  2.0$  respectively.  
For IMB, due to the higher energy threshold,  the
the dependence is linear in a restricted interval only. 

Due to their dependence on the difference of the original $\barnue$
and $\barnux$ fluxes, eq. (\ref{fluxe}), the conversion effects
increase with $r_E$.  In the wide energy range $E_{0e} = (8 - 11)$ MeV
(and for $r_L = 1$), both $\bar{\epsilon} ^{K2}$ and $\bar{\epsilon}
^{IMB}$ increase by up to $\sim 37\%$ with the variation of $r_E$
between 1 and 2.  This change is substantially larger than the
$1\sigma$ experimental interval, which equals $\sim 7.2-7.5\%$.  As
shown in fig. \ref{figure1.eps}, for $E_{0e}=11$ MeV the values of the
average energies are in the intervals $\bar{\epsilon}
^{K2}=17.5-23.5$ and $\bar{\epsilon} ^{IMB}=27.5-37.7$.

The dependence of $\bar{\epsilon}^{i}$ on $r_L$ is mild at
low-intermediate energies and becomes slightly stronger at high
$E_{0e}$ and for larger $r_E$.  For $r_E = 1.6$  we have that as $r_L$ varies from
0.67 to 1.5 the change in $\bar{\epsilon}^{i}$ does not exceed $\sim 7\%$ for K2 and $\sim 3.5\% $ for IMB.  
The latter dependence is weaker due to the higher IMB
energy threshold, which results in a reduced sensitivity of the IMB
signal to the softer component of the spectrum due to the original
$\barnue$ flux.

The plot (fig.~\ref{figure1.eps}) allows us to estimate the errors in
the determination of $E_{0e}$ in the
case when  conversion is not taken into account.  One can see that
in the no-oscillation case the observed average energy at K2
corresponds to $E_{0e} = 8.8$ MeV, while the IMB observation is
reproduced for $E_{0e}= 14.7$ MeV. With conversion these two values
become as low as $E_{0e} = 5.4$ MeV and $E_{0e} = 7.6$ MeV
respectively.

So, not taking into account conversion can lead
to an error in the determination of the  average energy (or temperature) of the original spectrum as large  
as 40 - 50\%.\\

2). The conversion leads in general to an increase of the number of
events (an exception is the case of low non-electron flux, $r_L < 1$,
and $r_E \sim 1$). The number of events is proportional to the
$\bar{\nu}_e$ luminosity, $L_e$: 
\beq N^{i} = N_0^{i} (E_{0e},
r_L, r_E) \frac{L_e}{L_0},
\label{scalen} 
\eeq where $N_0^{i}$ is the number of events in the detector $i$ for
a fixed luminosity $L_0$. For definiteness we take $L_0 = 5.3 \cdot
10^{53}$ ergs, which is the integral luminosity in the $\barnue$
species determined in ref. \cite{Loredo:2001rx} from the analysis of
data in the constant temperature scenario without oscillations.

In fig.~\ref{figure3.eps}  we show the dependences of $N_0^{i}$ for K2 and IMB
on $E_{0e}$ for different values of the ratios $r_L$ and $r_E$. Also shown are the
lines which correspond to the numbers of events observed  in these detectors
together with the $1\sigma$ bands.

While $N_0^{K2}$ increases nearly linearly with $E_{0e}$ (for
$E_{0e} > 6$ MeV), the dependence of $N_0^{IMB}$ is stronger as an
effect of the higher energy threshold and slow rise of the detection
efficiency with energy.

For $E_{0e} = 11$ MeV and $r_L = 1$ the increase of $N_0^{i}$ with
$r_E$ does not exceed $\sim 30\%$ for K2, while it can reach a factor
of 3.5 at IMB (see Table \ref{tab:2}). The increase is more sizable
for lower $E_{0e}$, as can be seen in fig.  \ref{figure3.eps}.

The numbers of events increase with $r_L$: we see a change by $\sim 30\%$  at K2 and by $\sim 75\%$ at IMB for  $E_{0e} = 11$ MeV and $r_E = 1.6$.

 In fig.~\ref{figure2.eps} we show the dependences of the total numbers
of events, $N ^{i}$, on $L_e$ for $E_{0e} = 11$ MeV and different
values of the ratios $r_L$ and $r_E$. This plot allows us to evaluate
the error in the determination of the original luminosity if the
oscillation effect is not taken into account. The detected number of
events $N^{K2} = 12$ at K2 can be reproduced without oscillations
for $L_e = L_0 = 5.3 \cdot 10^{52}$ ergs. With the increase of the
conversion effect (increase of $r_E$ or/and $r_L$) the required
luminosity decreases (see Table 1). E.g. we find: $L_e/L_0 = 0.78$ for
$r_E = 1.6$. For IMB the effect is much stronger: The observed $N
^{IMB} = 8$ events can be obtained for $L_e/L_0 = 2.75$ with no
oscillations. The required luminosity decreases down to $L_e/L_0 = 1.2$
for $r_E = 1.6$. \\

3). The conversion effects for K2 and IMB differ  due to

- different Earth matter effect; and

- different experimental characteristics: thresholds  and efficiencies
of detection.

In fig.~\ref{figure4.eps} we show the ratio of the numbers of events
at K2 and IMB, $r_N=N^{K2}/N^{IMB}$, as a function of $E_{0e}$. The
experimental value is $r_N = 1.5$. The ratio increases with
the decrease of $E_{0e}$. The conversion leads to a decrease of $r_N$,
thus allowing to reconcile the required energy $E_{0e}$ and the
correct value of $N^{K2}/N^{IMB}$. Taking $E_{0e} = 11$ MeV and $r_E = 1$
we find that $r_N$ varies between 4.2 and 1.5 depending
on $r_E$ (Table \ref{tab:1}).

In contrast  to  the numbers of events, the  energies of the observed events at K2 and IMB are modified 
by
the conversion in rather similar ways,  so that their ratio is not affected significantly: 
$\bar \epsilon^{K2}/\bar \epsilon^{IMB} = 0.62 -0.66$  (see Table \ref{tab:1} and 
fig.~\ref{figure3.eps}). The change  is even weaker for lower  $E_{0e}$.

The difference of conversion effects can be used to improve the agreement of the
IMB and K2 signals (sec.~\ref{sec4}).\\

4). In fig.~\ref{figure7.eps} we show the dependence of the widths of
the observed positron spectra, $\Gamma^{i}$, on $r_E$ for different
values of the pinching parameter $\alpha=\alpha_e=\alpha_x$ and
$r_L=1$ (results for other values of $r_L$ are almost identical). It
appears that this dependence is approximately linear for $r_E \gta
1.2$.  Conversion (i.e. $r_E > 1$) leads to widening of the K2 and IMB
observed spectra, and the widening is larger for larger $r_E$,
reaching $\sim 25 - 30\%$ for $r_E=2$. We see a $\sim 10\%$ decrease
of $\Gamma^{K2}$ and $\Gamma^{IMB}$ as $\alpha_e$ increases from 2 to
5 (i.e., as the original spectra become more pinched).

\begin{table}
\begin{center}
\begin{tabular}{lccccc}
\hline
scenario $\backslash$ $\alpha$     &  2    &  3       &  4 & 5  &  \\
\hline
\hline
no-oscillation ($\chi^2$)  &  24.05 &  16.8 &  17.0 & 19.2 \\ 
\hline
$N^{K2}$   & 13.0  & 11.9 & 11.2 & 10.7 & \\
\hline
$\bar \epsilon^{K2}/MeV$   &  19.4 &  17.5 &  16.4 & 15.5 & \\
\hline
$\Gamma^{K2}$   &  0.44  & 0.41   & 0.40   & 0.38  & \\
\hline
$N^{IMB}$   &  4.81 & 2.85 & 1.84 & 1.25 & \\
\hline
$\bar \epsilon^{IMB}/MeV$   & 30.1  &  27.5 &  25.9 & 24.8 & \\
\hline
$\Gamma^{IMB}$   &  0.33  & 0.31   & 0.30   & 0.29  & \\
\hline
\hline
 ``concordance'' ($\chi^2$)   &  15.0 &  11.0 &  12.4 & 14.7 & \\
\hline
$N^{K2}$   &  14.6 & 13.0 & 12.0 & 11.3  & \\
\hline
$\bar \epsilon^{K2}/MeV$   & 17.9 & 16.3 &  15.3 & 14.6 & \\
\hline
$\Gamma^{K2}$ & 0.47   &  0.45  & 0.44   & 0.43  &  \\
\hline
$N^{IMB}$   & 4.28  & 2.59 & 1.74 & 1.25 & \\
\hline
$\bar \epsilon^{IMB}/MeV$  & 30.7 & 28.5 & 27.0 &  26.0 &  \\
\hline
$\Gamma^{IMB}$    &  0.34  &  0.32  & 0.31   & 0.30  & \\
\hline
\hline

\end{tabular}
\caption{ Effects of  different values of $\alpha= \alpha_e =\alpha_x$ on the observables 
in the no-oscillation case with parameters from Loredo \& Lamb \cite{Loredo:2001rx} (lines 1 - 7) and in the case of 
oscillations with parameters from our concordance scenario given in secs. \ref{sec3} and \ref{sec4} (lines 8 - 14).}
\label{tab:2}
\end{center}
\end{table}

The dependence of the observables on $\alpha$ is summarized in the
Table \ref{tab:2}.  We show results for two scenarios: (1) the
no-oscillation best fit scenario of ref. \cite{Loredo:2001rx}, with
$E_{0e}=11$ MeV, $L_e=5.3 \cdot 10^{52}$~ ergs and (2) the
``concordance'' scenario (see sec. \ref{sec4}) with conversion,
$E_{0e}=8$ MeV, $L_e=8 \cdot 10^{52}$~ ergs, $r_E=1.6$ and $r_L=1$.

With the variation of $\alpha$, all the quantities in the table change
by at most $10\%$, with the exception of the number of IMB events,
which can vary even by a factor of 2 with respect to 
the case $\alpha = 3$.  In particular,  the decrease of $\alpha$ down to
$\alpha=2$, leads to an increase of $N^{IMB}$ by 65 - 70 \%.  That is, in
the ``concordance" scenario a smaller $\alpha$, $\alpha\simeq 2$, could
fit the numbers of events better.  This however is compensated by the
worsening of the fit to the average energies, so that the global fit
does not improve.


\section{SN1987A data and neutrino fluxes}
\label{sec4}

Using the results of the previous sections we now compare the data with
predictions and make qualitative statements on the interpretation of
experimental results and the determination of the original neutrino
fluxes.

Let us consider the integral characteristics of the signals. We assume
that all events are due to $\bar{\nu}_e$ interactions and
background. After background subtraction (according to
\cite{Hirata:1988ad,Bionta:1987qt}), we take the total numbers of
events detected by K2 and IMB as 
\beq N^{K2} = 12.0 \pm 3.5,
~~~~N^{IMB} = 8.0 \pm 2.8.
\label{numb} 
\eeq 
The experimental value of the average energy of the detected
events equals 
\beq 
\bar{\epsilon}^{k} = \frac{1}{N^{k}} \sum_i
\epsilon_i^{k}~, 
\label{ave} 
\eeq 
where the  summation runs over all observed
events in a given detector. We define the error of the average energy as 
\be
\Delta \bar{\epsilon}^{k} = \frac{1}{N^{k}} \sqrt{\sum_i (\Delta \epsilon^{k}_i)^2},
\label{err-av} 
\ee 
where $\Delta \epsilon^{k}_i$ is the error in the energy
determination of the individual event. 

Similarly, we calculate the width:
\be
\Gamma^{k} = \sqrt{N^{k} \frac{\sum_i (\epsilon^{k}_i)^2}{(\sum_i \epsilon^{k}_i)^2}-1}~,
\label{widcalc}
\ee 
and the associated error:
\be
\Delta \Gamma^{k} = \sqrt{\sum_i \left( \frac{\partial \Gamma^{k}}{\partial \epsilon^{k}_i} \Delta \epsilon^{k}_i \right)^2}~.
\label{generr}\ee

From the data we get the average energies:
\be
\bar \epsilon^{K2} = 14.7 \pm 1.1 ~{\rm MeV} ~~~~~~\bar \epsilon^{IMB} = 31.9 \pm 2.3~{\rm MeV},
\label{epsob}
\ee
and the widths:
\be
\Gamma^{K2} = 0.53 \pm 0.1,~~~~~~\Gamma^{IMB} = 0.23 \pm 0.06~.
\label{widexp}
\ee

\begin{figure}[hbt]
\begin{center}
\epsfig{file=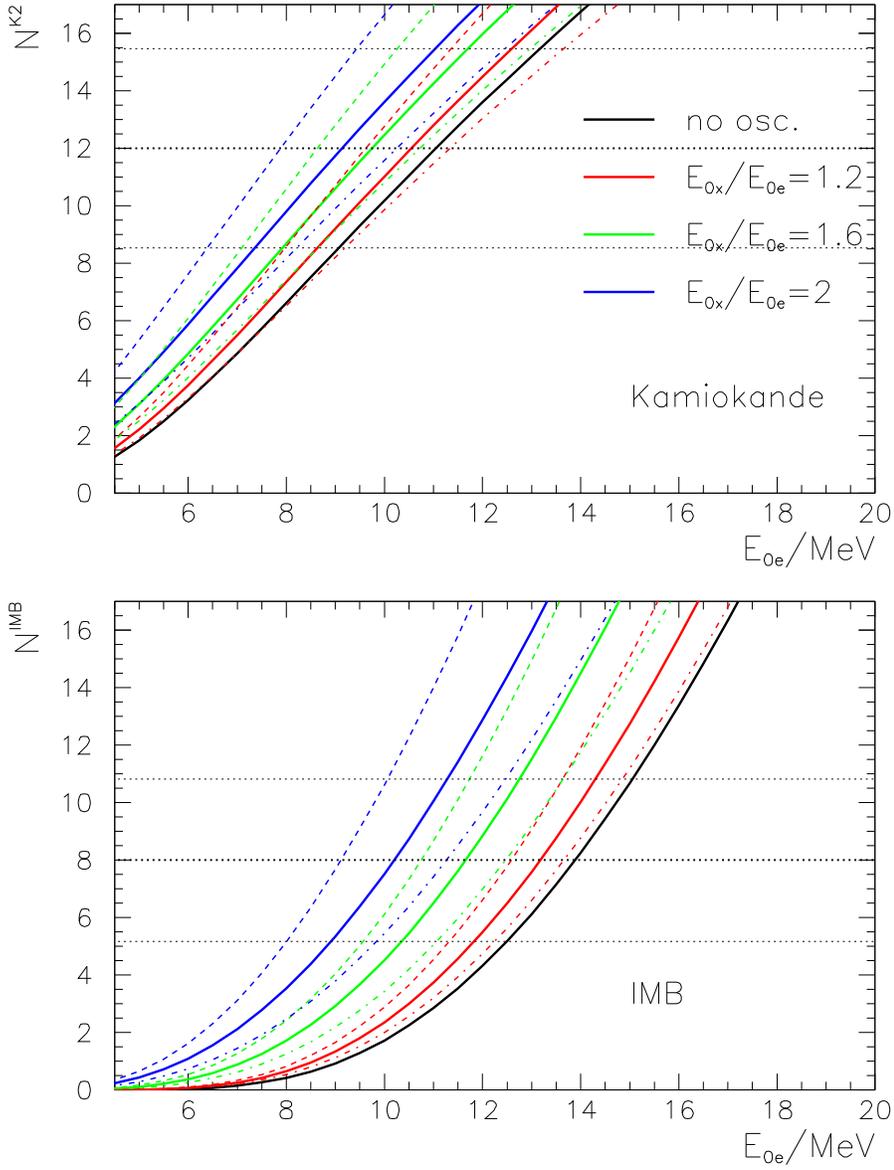, width=13truecm}
\end{center}
\caption{The predicted numbers of events at K2 and IMB as a function of the
average energy of the original $\bar \nu_e$ flux, $E_{0e}$. 
The line identification, as well as the oscillation parameters, are the same as  in fig.
\ref{figure1.eps}. 
The dotted (horizontal) lines represent the experimental
result with the $1\sigma$ error (eq. (\ref{numb})). We use $L_e=L_0 = 5.3 \cdot 10^{53}$ ergs.} \label{figure3.eps}
\end{figure}

\begin{figure}[hbt]
\begin{center}
\epsfig{file=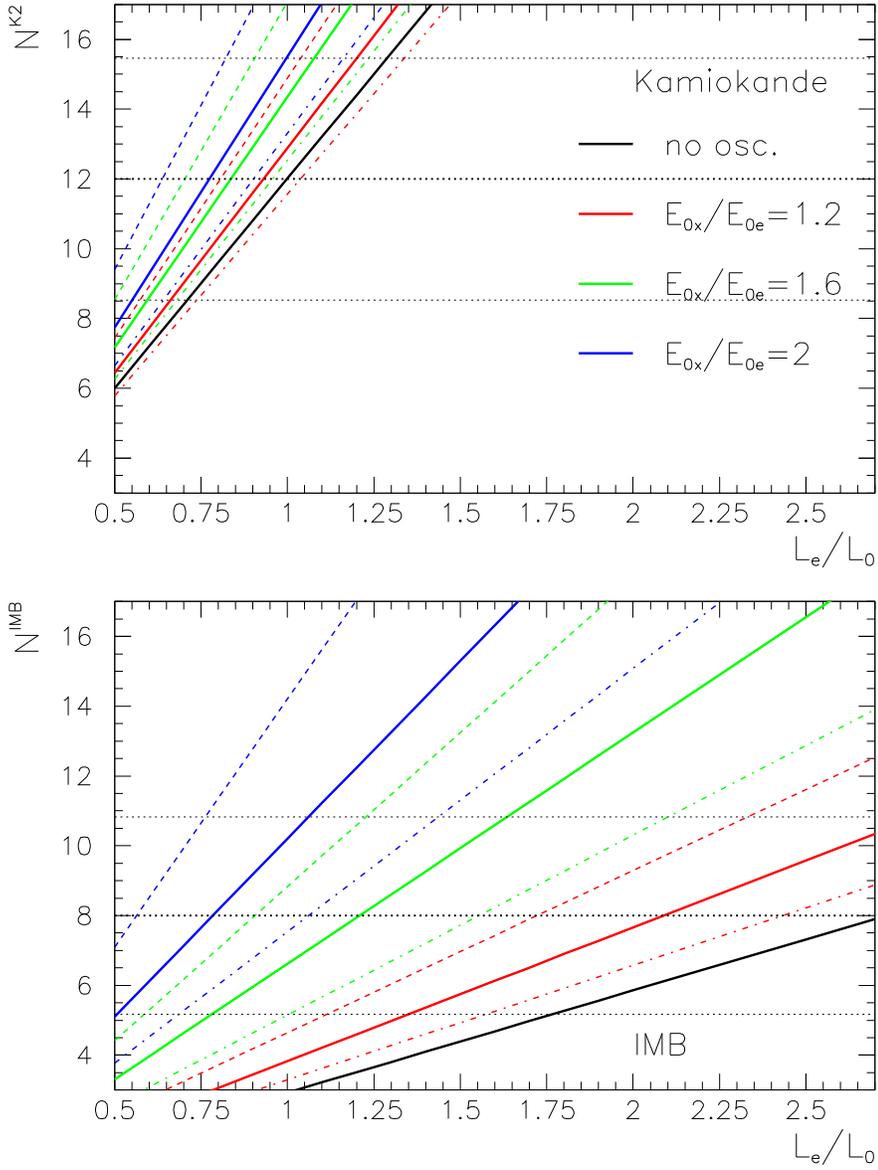, width=13truecm}
\end{center}
\caption{The predicted numbers of events at K2 and IMB as a function
of the
ratio $L_e/L_0$ ($L_0=5.3 \cdot 10^{52}$~ergs) for $E_{0e}=11$ MeV.
The line identification, as well as the oscillation parameters, are
the same as in fig.  \ref{figure1.eps}.
The dotted (horizontal)
lines represent the experimental result with the $1\sigma$ error (eq.
(\ref{numb})). } 
\label{figure2.eps}
\end{figure}

\begin{figure}[hbt]
\begin{center}
\epsfig{file=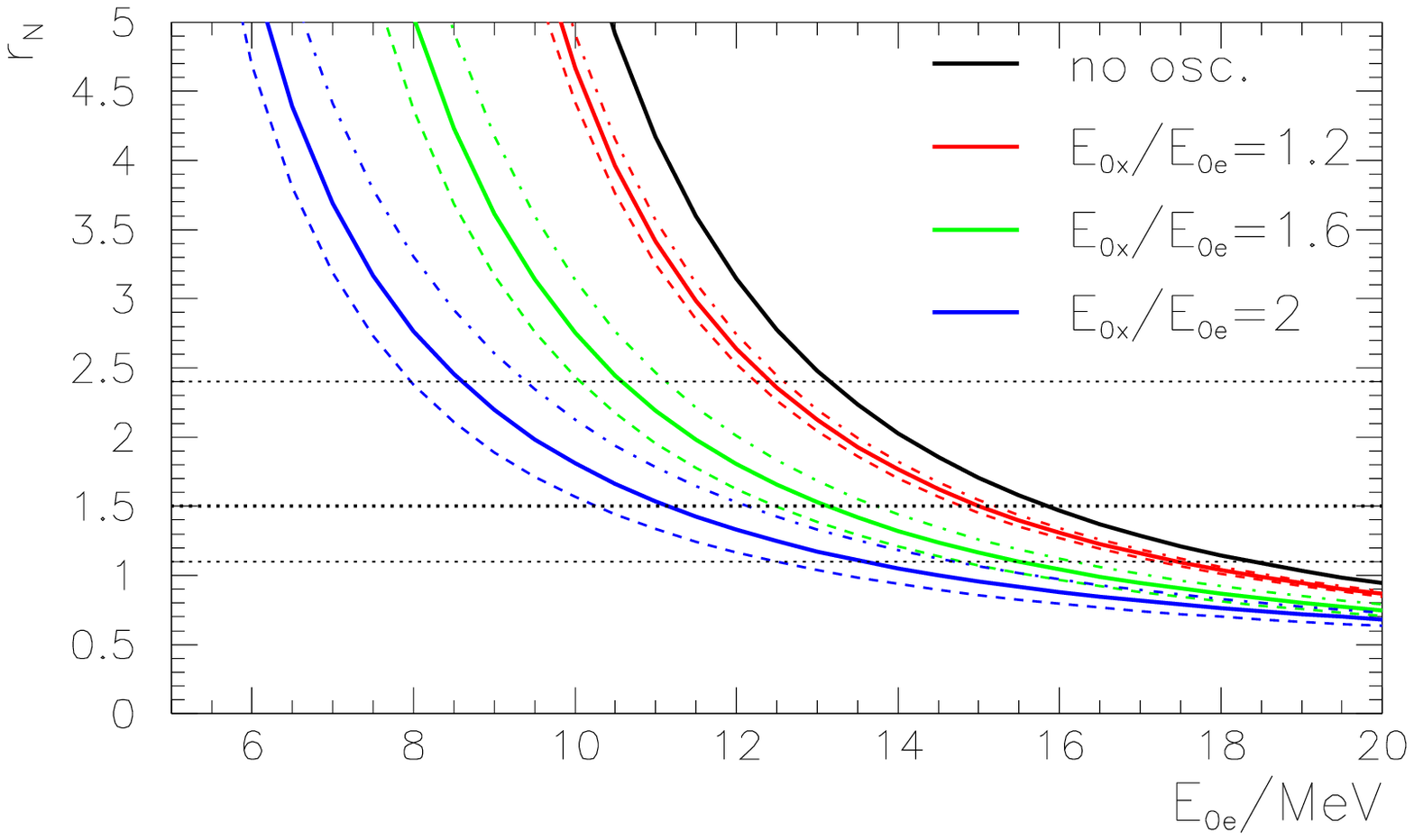, width=14.5truecm}
\end{center}
\caption{The ratio of the numbers of events at K2 and IMB, $r_N$, as a
a function of the average energy of the original $\bar \nu_e$ flux,
$E_{0e}$. 
The line identification, as well as the oscillation parameters, are the same as 
in fig. \ref{figure1.eps}.  The dotted 
(horizontal) lines represent the experimental result with the
$1\sigma$ error (eq. (\ref{rN})).}
\label{figure4.eps}
\end{figure}

\begin{figure}[hbt]
\begin{center}
\epsfig{file=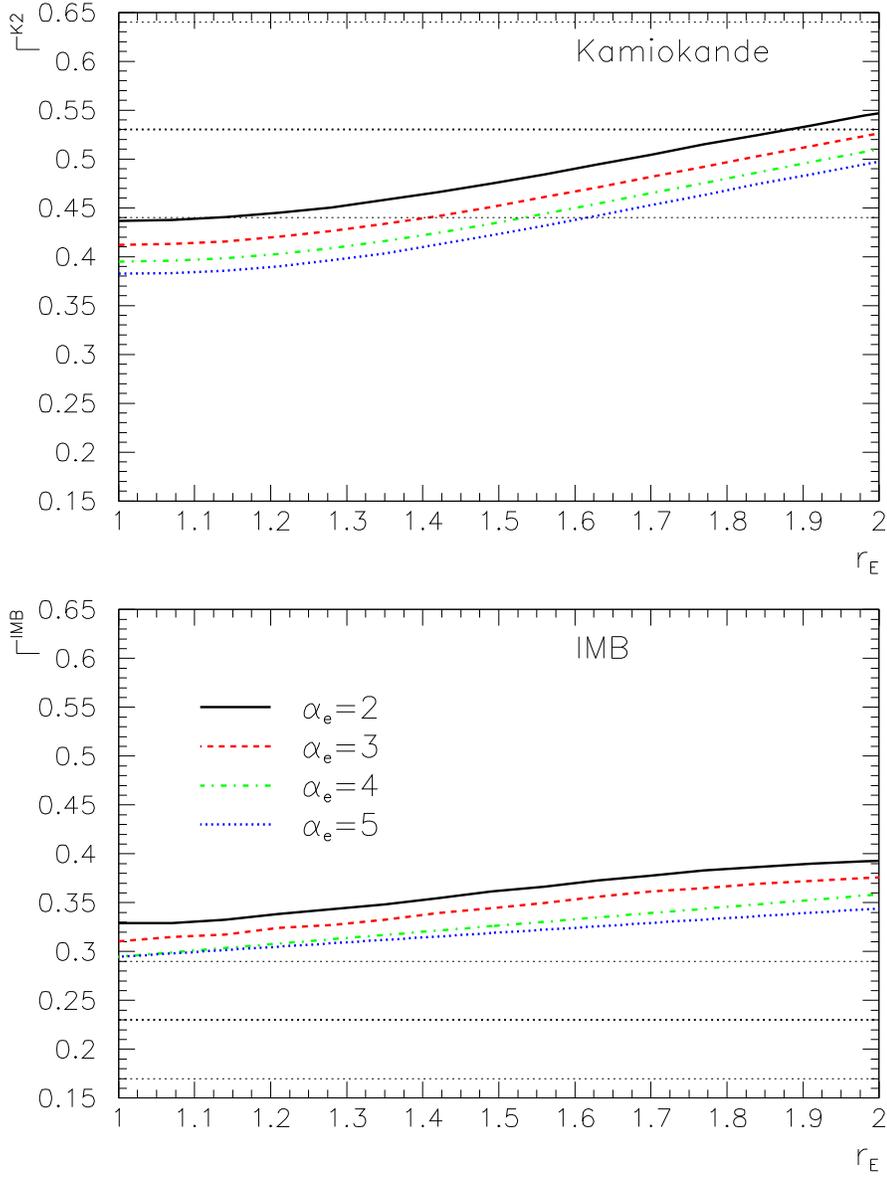, width=13truecm}
\end{center}
\caption{The predicted widths of the K2 and IMB energy spectra as a
function of $r_E$ for different values of the pinching parameter
$\alpha_e$.  The oscillation parameters are the same as in
fig. \ref{figure1.eps}. We have taken $E_{0e}=11$ MeV, $r_L=1$ and
$\alpha_x=\alpha_e$.  The dotted (horizontal) lines represent the
experimental result with the $1\sigma$ error, eq. (\ref{widexp}).}
\label{figure7.eps}
\end{figure}

We show these central values and the $1\sigma$  bands 
in the figs. \ref{figure1.eps} - \ref{figure7.eps}.

To compare those results with observations (\ref{numb}, \ref{epsob})
one needs to perform an integration over time, taking into account the
time dependence of the parameters of the original spectra
(alternatively, one could analyze the data in short periods of time,
however the statistics is too small).  Indeed, it is expected that
during the neutrino burst there is a significant change in the
parameters of spectra with time~(see e.g. \cite{Totani:1998vj} and
references therein).  However, due to the nearly linear dependences of the
observables on the parameters, the relative conversion effect changes
with time very weakly.

Let us perform time averaging of the relation (\ref{avE}). If the
ratios $r_E$ and $r_L$ depend on time weakly -- which is expected
during the cooling phase -- then the $A$ and $B$ coefficients are nearly
constant and for the time averaged energy we can write 
\be 
\langle
\bar{\epsilon} \rangle \approx A(\bar r_E, \bar r_L) \langle E_{0e} \rangle
+ B(\bar r_E, \bar r_L), \label{avEt} 
\ee 
where $\langle E_{0e} \rangle = \int
dt E_{0e} / \Delta t$. This means that the relation (\ref{avE}), and
the dependences in fig.~\ref{figure3.eps} can be considered as the
relation and dependences between the time averaged characteristics.

Let us now consider numbers of events.
Assuming  linear dependence of $N_0^{i}$ on
$E_{0e}$, $N_0^{i} \approx C^{i}(r_E, r_L) E_{0e} + G^{i}(r_E,
r_L)$ (which may not be a good approximation for IMB), and weak change
or $r_E$ and $r_L$ with time, we can write: 
\beq 
N^{tot,i} = \left[
C^{i}(\bar{r}_E, \bar{r}_L) \langle {E}_{0e}\rangle_L +
G^{i}(\bar{r}_E, \bar{r}_L)\right] \frac{1}{L_0} \int dt L(t) , \eeq
where \beq \langle {E}_{0e} \rangle_L = \frac{\int E_{0e} L dt}{\int L dt}. 
\label{eL} 
\eeq

From this it follows that we can use the relations and figures
constructed for instantaneous spectra or -- equivalently -- the
time-independent scenario, also for the case of time-dependence,
keeping in mind that the parameters $E_{0e}$, $L_e$, $r_E$, $r_L$
should be considered as some effective parameters obtained by
appropriate averaging over the time of the burst. This still can be
used to study the compatibility of signals in different detectors, but
the real meaning of the parameters extracted is ambiguous and it
depends on the detailed time dependence of instantaneous
characteristics.  In what follows we will use the same notations as in
sec. 3, though one should keep in mind that these are  effective
parameters. \\

1). {\it Analysis of the Kamiokande-2 data only}. In the absence of
oscillations, from fig.~\ref{figure1.eps} we extract the value
of the (effective) $\bar \nu_e$ average energy: 
\beq 
E_{0e} = 8.7
  \pm 0.9 ~{\rm MeV}, ~~~ (1\sigma). 
\label{e-kam} 
\eeq  
According to
fig.~\ref{figure3.eps}, for this energy and $L_e = L_0$ the number of
predicted events equals $N_e = 7.8 \pm 1.5$ which is about $1\sigma$
below of observed number, eq.  (\ref{numb}). The exact number of
observed events can be reproduced if $L_e = 1.54 L_0 = 8.2 \cdot
10^{52}$ ergs. 

The conversion leads to a decrease of $E_{0e}$. For instance, for $r_E
= 1.6$ and $r_L = 1$ we find from fig.~\ref{figure1.eps}: 
\beq E_{0e}
= 6.8 \pm 0.8 ~{\rm MeV}, ~~~ (1\sigma).  
\eeq 
According to
fig.~\ref{figure3.eps}, this gives $N ^{K2} = 6.3 \pm 1.5$ for $L_e =
L_0$. So, correspondingly, a fit to the observed number of events
implies even higher integral luminosity: $L_e = 1.9 L_0 = (10.1 \pm
2.4) \cdot 10^{52}$ ergs, which seems rather high, also keeping in
mind the low average energies of neutrinos.

So, the K2 data alone disfavor strong conversion effect; though within
$2 \sigma$ even a strong effect ($r_E = 1.6$ ) can be easily
accommodated. \\

2). {\it Kamiokande-2 and  IMB}. For IMB, in the absence of
oscillations, we find, from fig.~\ref{figure1.eps}: 
\beq 
E_{0e} = 14.7 \pm 1.9 ~{\rm MeV}, ~~~ (1\sigma)  
\label{e-imb} 
\eeq 
which is more than $3\sigma$ above the K2 result, eq. (\ref{e-kam}). According to
fig.~\ref{figure3.eps} for this energy and $L_e = L_0$ the number of
predicted events equals $N^{IMB} = 10 +6/-4$ which is in a very good
agreement with the experimental result. The precise number of the IMB
events can be reproduced for $L_e = 0.8 L_0$.

Thus, 
the IMB signal implies about 2
times higher energies of events and 2 times smaller integral luminosity in
comparison with K2. 

Conversion leads to a decrease of the extracted energy and
luminosity. For $r_E = 1.6$ and $r_L$ we get \beq E_{0e}^{IMB} =
10.3 \pm 1.7 ~{\rm MeV}, ~~~ (1\sigma).
\label{e-imb2} \eeq which is only $2\sigma$ above the K2 result for the same
$r_E$.

With conversion effects, the agreement of the IMB and
K2 results can improve. We can find a ``concordance" set of
parameters of the original fluxes which give a better description of
all the available data: average energies, luminosities and widths. The
latter can be affected significantly by the oscillations in the matter
of the Earth.

The ratio of the average energies extracted from the K2 and IMB data
approaches 1 with the increase of $r_E$: $E_{0e}^{IMB}/E_{0e}^{K2} =
1.7, 1.51, 1.43$ for $r_E = 1.0, 1.6, 2.0$. Notice however, that the
difference does change significantly for $r_E > 1.6$ being about
$2\sigma$. Furthermore, with the increase of $r_E$ both the extracted
energies $E^{K2}_{0e}$ and $E_{0e}^{IMB}$ decrease, and this makes
it difficult to reproduce the total numbers of events, especially for
IMB. Considering this, we find that the optimal energy is in the
interval $E_{0e} = (7 - 9)$ MeV, where one can get $1\sigma$ deviation
from both K2 and IMB extracted values. This result does not depend on
the luminosity $L_e$.

Considering the numbers of events, notice that once the correct ratio
$r_N \equiv N ^{K2}/N ^{IMB}$ is reproduced, the numbers of events can
be obtained by fitting $L_e$. From the data we find 
\beq r_N \equiv
\frac{N ^{K2}}{N ^{IMB}} = 1.5^{+0.9}_{-0.4}.
\label{rN} 
\eeq 
However, according to fig.~\ref{figure4.eps}, for the small $E_{0e}$
required by the fit of the average energies the ratio $r_N$
turns out to be too large. Indeed, for $r_E = 1$ ({\it i.e.}, no
oscillations), the ``optimal'' energy is $E_{0e} = 10.3$ MeV; for
these parameters the ratio of number of events equals $r_N = 5.3$. We
find $E_{0e} \sim 9.6$ MeV and $r_N = 5.2$ for $r_E = 1.2$; $E_{0e}
\sim 7.7$ MeV and $r_N = 5.2$ for $r_E = 1.6$; $E_{0e} \sim 6.2$ MeV
and $r_N = 4.9$ for $r_E = 2.0$.  So, $r_N$ is substantially larger
than the experimental result and does not change practically by
conversion once the parameters are optimized by the average observed
energies.

Let us now discuss the widths of the observed spectra $\Gamma^{i}$,
shown in fig. \ref{figure7.eps} (for $E_{0e}=11$ MeV and $r_L=1$). One
can see (see also eq. (\ref{widexp})), that the IMB spectrum has
smaller width than the one of K2. The former is better reproduced
without oscillations (or small $r_E$) and with large $\alpha$,
$\alpha\simeq 5$, while the latter points toward the case of
oscillations with $r_E\gta 1.6$ and $\alpha\sim 2$. The widths
$\Gamma^{i}$ decrease slightly (by $\sim 10\%$) with the decrease of
$E_{0e}$; we refer to Table \ref{tab:2} for results at
$E_{0e}=8$ MeV (concordance scenario).  \\


To get an idea about the preferable values of parameters of the original spectra
we calculate $\chi^2$  defined as 
\be 
\chi^2 \equiv \sum_j \left(\frac{X_j -
\bar{X}_j}{\Delta X_j} \right)^2 
\ee 
for various sets of the parameters. Here
$X_j \equiv N ^{K2}, N ^{IMB},\bar \epsilon ^{K2}, \bar \epsilon ^{IMB}, \Gamma^{K2}, \Gamma^{IMB}$,  are
the numbers of events, average energies and widths of the positron spectra observed at Kamiokande2 and IMB.

Without oscillations and $E_{0e} = 11$ MeV, $L = L_0$,
$\alpha_e=\alpha_x=3$ we find $\chi^2 = 16.2$. With oscillations for
$E_{0e} = 8$ MeV, $L = 8 \cdot 10^{52}$ ergs, $\alpha_e=\alpha_x=3$,
$r_E = 1.6$, $r_L = 1$, we obtain $\chi^2 = 11.0$ (see Table
\ref{tab:2} for the dependence of results on the pinching parameters
$\alpha_z$).
We
checked that the improvement in the $\chi^2$ is mainly due to the smaller
deviation of the predicted $\bar \epsilon ^{K2}$ from the observed
value. This deviation amounts to about 2.6 $\sigma$ for
no-oscillations, and to $\sim 1.4~~ \sigma$ for the case with
oscillations. Correspondingly, the contribution of $\bar \epsilon
^{K2}$ to the $\chi^2$ decreases from 7 to 2 when oscillations are
included.  Notice however that with the decrease of $E_{0e}$ implied by
conversion effects, substantially larger integral luminosity is
required to reproduce the numbers of events.
The combination of smaller average energy and larger luminosity
corresponds to a larger radius of the neutrinosphere: $R_{ns}\propto
E_{0e}^{-2} L^{1/2}$. It follows that in our concordance scenario
$R_{ns}$ is about 2.4 times larger than in the no-oscillation
scenario, which gives $R_{ns} \simeq 20-30$ Km \cite{Loredo:2001rx}.
This increase is within the range allowed by the several uncertainties
of astrophysical nature and by individual variations from one star to
another.

Thus we conclude that 
conversion leads to a certain improvement of the global fit of the data.  
The improvement  requires lower average energies of the original  $\bar{\nu}_e$ spectrum
and  higher $\bar{\nu}_e$ integral luminosity.  
Conversion effects lead to a better agreement of the observed
average energies with the data sample, but this improvement is 
compensated partially by a worsening in reproducing the numbers of events.

3). {\it Baksan results.} If all the 5 events recorded by Baksan are
attributed to signal, analyses  of these data without oscillations  lead to
neutrino luminosities which are much larger than those implied by the K2 and IMB
data samples (see the account in ref. \cite{Loredo:2001rx}).  This indicates 
that part of the Baksan events are due to background, and, in absence of a
specific prescription for background subtraction, it is difficult 
to include the Baksan results in our analysis. 
In ref. \cite{Loredo:2001rx} a method to account for a background origin of part of the
Baksan events is proposed. 
Notice that in presence of oscillations the K2 and IMB data require
larger neutrino luminosities, therefore it is possible that including
neutrino conversion will reduce the number of the Baksan
events attributed to background.\\

4). {\it Spectrum of events.} As we have marked in the introduction,
the oscillation parameters we use are in the band determined
previously from the fit of the energy spectra of K2 and IMB
\cite{Lunardini:2000sw} (see also fig. \ref{fig:lma}). So the
oscillation parameters, and in particular $\Delta m^2$, are already
optimized to get the best description of the spectra of events.  We mark
that, in contrast with what discussed in \cite{Lunardini:2000sw}, here
the numbers of events, widths  and average energies of the observed positrons
depend very weakly (about $\sim 2-5\%$ at most) on the phase of
oscillations in the Earth. This is motivated by the larger value of
$\Delta m^2$ used here, for which the permutation factor has faster
oscillations with smaller depth. These oscillations can be seen in
figure \ref{fig:lma} (in which no averaging has been applied), however
in the observed signal they are almost completely averaged out as an
effect of the experimental energy resolution.

\begin{figure}[hbt]
\begin{center}
\epsfig{file=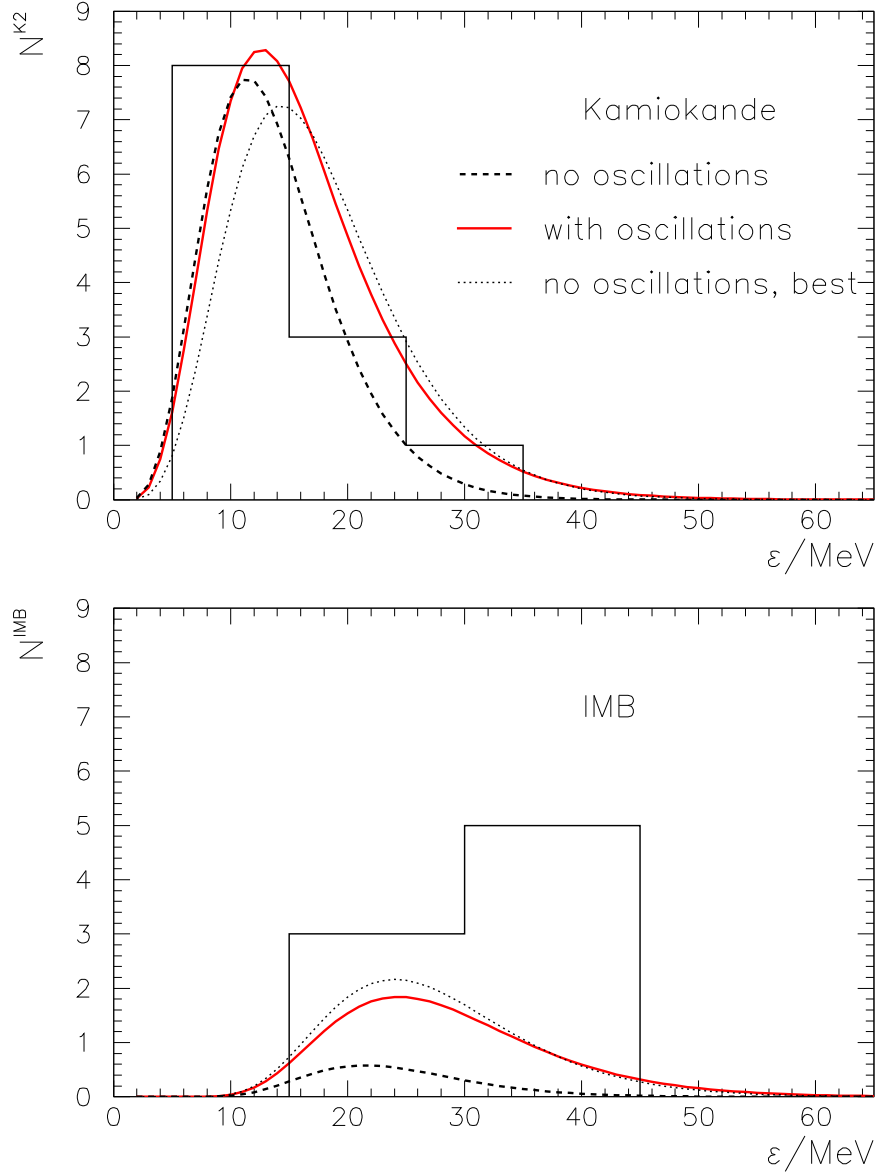, width=13truecm}
\end{center}
\caption{The predicted positron energy spectra at
K2 and IMB with and without oscillations. The solid 
lines  correspond to  $E_{0e} = 8$ MeV, $L_e = 8 \cdot 10^{52}$ ergs, $r_E
= 1.6$, $r_L = 1$ and the best-fit LMA parameters (eq. (\ref{bfglob}),
see also fig. \ref{fig:1}). 
The dashed lines correspond to the no-oscillation case with the 
same parameters: $E_{0e} = 8$ MeV, and $L_e = 8 \cdot 10^{52}$ ergs. 
The dotted lines correspond to
no-oscillations with the best fit parameters of ref.
\cite{Loredo:2001rx}: $E_{0e} = 11$ MeV, $L_e = 5.3 \cdot 10^{52}$
ergs.  The histograms show the experimental results. } \label{figure5.eps}
\end{figure}

In fig.~\ref{figure5.eps} we show the calculated spectra of the events
at Kamiokande-2 and IMB for the ``concordance'' set of parameters of the
original fluxes: $E_{0e} = 8$ MeV, $L_e = 8 \cdot 10^{52}$ ergs, $r_E
= 1.6$, $r_L = 1$ (solid lines). Also shown are the spectra without
oscillation for the best fit parameter set determined in
\cite{Loredo:2001rx} (``constant temperature model") (dotted) and the
spectrum for the concordance set of spectral parameters without
oscillations (dashed).  Comparing solid and dashed lines we conclude
that the conversion effect is significant in the whole energy
range  for IMB and  for $\epsilon > 15$ MeV at K2.

At the same time the spectra with best fit  values of parameters (without
conversion) and with the concordance set of parameters and conversion are
rather close. The difference is mainly in the low energy K2 spectrum.
We stress that, though similar,the spectra with and without conversion
require substantially different parameters of the original spectra.

\section{Discussion and Conclusions}

1). After the identification of the solution of the solar neutrino
problem and KamLAND results, we can definitely say that neutrinos from
SN1987A underwent  flavor conversion inside the star and
oscillations in the matter of the Earth.

In the assumption of the normal mass hierarchy, the conversion
probabilities can be calculated with good precision.
We find that the permutation factor is about $p= 0.28 - 0.32$ due to conversion inside the star  
and oscillations in the matter of the  Earth 
suppress the permutation. The Earth effect increases with
energy, and   at $E \sim 40$ MeV $p$ decreases down to 0.15- 0.20.\\

\noindent
2). The conversion effects on observables depend strongly on the
properties of the original neutrino fluxes, in particular, on the average
energies and widths of the original spectra of $\bar{\nu}_e$ and
$\bar{\nu}_{\mu/\tau}$.
For a given set of these parameters, conversion leads to an
increase of the number and of the average energy of the observed events
as well as of the widths of the observed spectra. The conversion effects
are different for K2 and IMB. By varying parameters in the ranges
allowed by astrophysics, we have found that the average energy of
events can increase by 30 - 50 \% and the number of events by 50 \%
for K2 and by up to a factor of 3 for IMB.

{\it Vice versa}, conversion changes the average energies and
luminosities of the original neutrino fluxes extracted from the
observations. In particular, it leads to a decrease of the original
energy and (for fixed $E_{0e}$) of the luminosity of $\bar{\nu}_e$. We find that the decrease
can be up to $\sim 50-70\%$. This leads to uncertainty in the
determination of parameters of the original spectra.\\

\noindent
3). Comparing calculations with the real signals from SN1987A we find
that the K2 data alone do not show significant conversion effects, and
can be well described by original fluxes with small conversion
effects. This would testify for small difference of the original
$\bar{\nu}_e$ and $\bar{\nu}_{\mu}$ spectra. At the same time the K2
data do not exclude strong conversion. In this case, however, the
original spectra should have lower energies and higher luminosities in
comparison with the no oscillation case.

The characteristics of the original spectra extracted from the K2 and
IMB data exhibit substantial differences.  These can be partially
reduced by conversion effects: that is, the conversion improves the
combined fit of the K2 and IMB data. The improvement is not
dramatic, though, and does not lead to substantially more coherent overall
picture.  It requires strongly different original spectra, $r_E = 1.5
- 2$, low average energy of the $\bar{\nu}_e$-spectrum, $E_{0e}=6.5 -
8.5$ MeV, and high luminosity: $L_e = (8 - 12) \cdot 10^{52}$ ergs.

The oscillations in the matter of the Earth substantially modify 
the high energy parts of the spectra of events at K2 and IMB.

Our conclusions are valid for normal mass hierarchy or inverted
hierarchy provided that $\theta_{13}$ is negligibly small.  If the
hierarchy is inverted and $\sin^2 \theta_{13}\gta few \cdot 10^{-4} $,
the $\barnue \leftrightarrow \barnux$ permutation in the star is complete
\cite{Dighe:1999bi,Lunardini:2003eh}, so that the $\barnue$ flux at
Earth is entirely due to the original $\numu/\nutau$ flux. It follows that 
the parameters extracted from the data analysis refer to the
non-electron flavors produced inside the star and therefore would lead
to a completely different test of supernova theory, with respect to
the non-oscillation case.

In this same scenario of mixing and hierarchy, the amount of $\barnue
\leftrightarrow \barnux$ permutation could change at late times, as the
supernova shock-wave reaches the external layers of the star, thus
modifying the adiabaticity character of the $\theta_{13}$-induced MSW
resonance
\cite{Schirato:2002tg,Takahashi:2002yj,Lunardini:2003eh,Fogli:2003dw}.
The effect on the time integrated neutrino signal is however small
\cite{Takahashi:2002yj,Fogli:2003dw} and negligible with respect to
the large uncertainties of statistical and astrophysical nature on the
SN1987A data.\\

\noindent
4). In general, one needs to perform an analysis which employs the
whole information contained in the data: both integral and 
differential (energy spectrum, arrival time) as well as errors in the
determination of the energies of events, background and angular
information. In view of the small number of detected events, the optimal
type of analysis would be along the lines of the work of Loredo and Lamb
\cite{Loredo:2001rx}, with the conversion effects taken into
account. The present study will allow to better understand the results
of such a global fit. Our conclusions concerning the interpretation of
the observed signals have qualitative character only.\\

\noindent
5). It would be important to compare these results on SN1987A with
those of future SN neutrino detections.  The latter will have high
statistics and therefore will provide the possibility to disentangle
the oscillation effects and the properties of the original fluxes.  We
mark that many of the results presented here have general character
and therefore will apply to future data as well. In particular, the
effects of conversion in the star will be valid.  At the level of
probabilities (permutation factor), the effects of oscillations in the
Earth will probably be different, due to different trajectory of the
neutrinos in the Earth. However the difference with respect to SN1987A may be
negligible in the observed energy spectra if the energy resolution of
the detector is larger than the size of the spectral modulations due
to oscillations.
The analysis of future supernova data will result in a better
understanding of the generation of neutrino fluxes and therefore in
more reliable predictions of fluxes also for SN1987A. In this way a
more precise interpretation of the SN1987A signals can be done.

\subsection*{Acknowledgments}

C.L. acknowledges support from the Keck fellowship at IAS and the NSF grant
PHY-0070928. She would like to thank C. Pe\~na-Garay and L. Sorbo for useful
discussions. 

\subsection*{Note added}
It has been proposed recently that the signal detected by the LSD detector
at Mont Blanc \cite{Dadykin:ek}, 4.7 hours before the Kamioka-2 and IMB
signals, is due to the primary collapse of a fastly rotating Fe-O-C
stellar core that leads to the formation of a collapsar
\cite{Imshennik:2004iy}. The latter undergoes fragmentation into a
binary of neutron stars, which loose angular momentum by emission of
gravitational waves.  This model predicts an early neutrino burst,
produced in the URCA processes (neutronization), and a second one,
after several hours, which results from the collapse of the heavier
star of the binary system.

The early burst 
is composed
dominantly of  $\nu_e$'s with average energies of 30 - 40 MeV.  The
signal in LSD is mainly due to CC interactions of these $\nu_e$'s with
nuclei of Fe in the walls of the detector. The produced electrons
generate  electromagnetic showers which loose energy in the walls
and appear in the scintillator as signals with energies of 7 - 10
MeV.
                                                                                
Neutrino conversion can substantially change this picture.  In the
most plausible scenario of normal hierarchy and $\tan^2\theta_{13} \gta
10^{-4}$ the $\nu_e$ flux is  almost completely converted to
a $\nu_{\mu}/\nu_{\tau}$ flux in the star
\cite{Dighe:1999bi,Lunardini:2003eh}. Furthermore, the regeneration
effect in the Earth is negligible. So, the proposed signal due to the
CC- interactions in the LSD detector should not appear.

In the case of very small 1-3 mixing, $\tan^2\theta_{13} \ll 10^{-4}$
or inverted mass hierarchy, the signal
is  suppressed by the factor $\sin^2 \theta_{12} \sim 0.3$
(though some regeneration effect in the matter of the Earth is
present). In this case the explanation of the LSD
signal would require  an original neutrino luminosity about 3 times larger 
than that in \cite{Imshennik:2004iy}.
                                                                                
We mark that the LDS signal could be explained by NC
 interactions induced by the  $\nu_{\mu}/\nu_{\tau}$ flux. However, due to the small
 cross section, to reproduce the observed number of events a much
 larger original luminosity would be necessary.

\bibliography{noficeci}


\end{document}